\documentclass[12pt,twoside]{article}
\usepackage[mathscr]{eucal}
\usepackage{amsmath,amsfonts,amssymb,amsthm,mathabx,empheq}
\usepackage{times}
\usepackage{pdfsync}
\usepackage{cite}
\usepackage{url}
\usepackage{hyperref}
\usepackage{tensor}
\usepackage{color}
\usepackage{multicol}
\usepackage{bbold}

\advance\textheight\topskip
\allowdisplaybreaks[1]

\newcommand\td{\text{d}}

\newcommand{\p}{\partial}

\newcommand{\be}{\begin{equation}}
\newcommand{\ee}{\end{equation}}
\newcommand{\bea}{\begin{eqnarray}}
\newcommand{\eea}{\end{eqnarray}}
\def\nn{\nonumber}
\def\bz{\bar z}

\def\cY{{\cal Y}}
\def\bY{\bar Y}

\def\bP{\bar P}

\def\bL{\bar{L}}

\def\bomega{\bar\omega}
\def\bY{\bar Y}

\def \sd {\delta\hspace{-0.50em}\slash\hspace{-0.05em}}

\newcommand*\xbar[1]{%
  \hbox{%
    \vbox{%
      \hrule height 0.5pt 
      \kern0.3ex
      \hbox{%
        \kern-0.0em
        \ensuremath{#1}%
        \kern-0.0em
      }%
    }%
  }%
}

\DeclareFontFamily{OT1}{rsfs}{} \DeclareFontShape{OT1}{rsfs}{m}{n}{
<-7> rsfs5 <7-10> rsfs7 <10-> rsfs10}{}
\DeclareMathAlphabet{\mycal}{OT1}{rsfs}{m}{n}

\begin{document}
\title{Null boundary gravitational charges from local Lorentz symmetries}

\author{Pujian Mao and Weicheng Zhao}

\date{}

\def\mytitle{Null boundary gravitational charges from local Lorentz symmetries}

\addtolength{\headsep}{4pt}

\begin{centering}

  \vspace{1cm}

  \textbf{\Large{\mytitle}}

  \vspace{1.5cm}

  {\large  Pujian Mao and Weicheng Zhao}

\vspace{.5cm}
\begin{minipage}{.9\textwidth}\small \it  \begin{center}
     Center for Joint Quantum Studies and Department of Physics,\\
     School of Science, Tianjin University, 135 Yaguan Road, Tianjin 300350, China
 \end{center}
\end{minipage}

\end{centering}


\vspace{0.1cm}

\begin{center}
\begin{minipage}{.9\textwidth}
\textsc{Abstract}. In this paper, we revisit the null boundary gravitational charge in the Newman-Penrose formalism with special emphasis on the charges from local Lorentz transformations. We find that there is one more charge derived from the local Lorentz transformation and the new charge is purely from the Holst term. This reveals a remarkable fact that trivial terms which do not change classical equations of motion can not only affect the boundary degrees of freedom through their contributions to the boundary charges but also have their own rights to create new boundary degrees of freedom.
 \end{minipage}
\end{center}
\thispagestyle{empty}


\section{Introduction}

When spacetime equipped with a boundary, new degrees of freedom which reside on the boundary should be added to the system of relevance. The boundary degrees of freedom are labelled by boundary (surface) charges. The boundary charges are of importance for gravitational theories. Because the energy or mass of gravitational theory is defined as boundary charge at spatial \cite{Arnowitt:1962hi} or null \cite{Bondi:1962px} infinity. Recently, there have been renewed interests on first order description of gravitational theory, in which dual mass can be formulated as boundary charges from the Holst term that added to the Palatini action \cite{Godazgar:2020gqd,Godazgar:2020kqd}. Strictly speaking, the derivation in \cite{Godazgar:2020gqd,Godazgar:2020kqd} just reveals that the Holst term does contribute to the boundary charge and its contribution is the dual mass studied previously in \cite{Godazgar:2018qpq}, see also \cite{Godazgar:2018vmm,Kol:2019nkc,Godazgar:2019dkh,Oliveri:2020xls}. In first order formalism, the gauge symmetries of the gravitational theory consist of both diffeomorphism and local Lorentz transformation \cite{Julia:1998ys,Julia:2000er,Ashtekar:2008jw,Corichi:2013zza,Jacobson:2015uqa,Barnich:2016rwk,DePaoli:2018erh,Barnich:2019vzx,Oliveri:2019gvm,Barnich:2020ciy}. It is shown that the local Lorentz transformation is trivial in the sense that it does not have associated charge from the Palatini action nor the Holst term \cite{Godazgar:2020kqd}. Hence, the Holst term will not arise any new boundary charge based the current  investigations. Then, if the Holst term can arise
new boundary degree of freedom is not clear and it is precisely what we will address in the present work.

In this paper, we study near horizon boundary charges in the context of Newman-Penrose (NP) formalism. By relaxing the boundary conditions chosen in \cite{Liu:2022uox}, we find that the near horizon symmetry consists of near horizon supertranslations and superrotations from the horizon diffeomorphism and two independent Lorentz transformations. One of the Lorentz transformation is equivalent to the Weyl scaling of the horizon discussed in \cite{Adami:2021nnf}. The other is a complex rotation of the null basis which generates completely new near horizon symmetry. We work out the boundary charges. Interestingly, the boundary charge of the complex rotation is only from the Holst term. So the complex rotation is a trivial gauge transformation in the usual sense in the theory determined by the Palatini action. This is a remarkable example in the sense that trivial residual gauge symmetry can be made large when trivial terms are added to the action. In total, there can be five charges defined on a three dimensional null hypersurface. The charge algebras for the Holst term are also derived. There is a non-trivial 2-cocycle term between the near horizon supertranslation and the complex rotation. The 2-cocycle term is field dependent which depends on the area of the horizon. If the boundary charges are related to physical observables, our derivations, the new charge and charge algebras, reveal the observational effect from trivial terms. This can serve as a criterion to fix the trivial or boundary terms in the action. It is worthwhile to emphasize that the new perspective of the present work is not about the Lorentz charges but to show that Holst term will arise new boundary degree of freedom, namely new boundary charges can not be seen from the Palatini action. Relevant studies on Lorentz charges can be seen, for instance, in \cite{Jacobson:2015uqa,Freidel:2020xyx,Freidel:2020svx,Freidel:2021fxf,Godazgar:2022foc}.

This paper is organized as follows. In the next section, we specify our gauge and boundary conditions and work out the corresponding residual gauge transformations, near horizon solution space and transformation laws of the near horizon fields. In section \ref{Palatinisection}, we derive the boundary charges from the Palatini action. The main point of this section is to show that there is no charge associated to the new near horizon symmetry, i.e., the complex rotation. In section \ref{Holstsection}, we compute the boundary charges and charge algebras from the Holst term. The last section is devoted to discussions and outlooks of our results.

\section{Null boundary symmetries and solution space}
\label{solution}

In this section we will work out the null boundary symmetries and solution space in the NP formalism \cite{Newman:1961qr}. We follow precisely \cite{Liu:2022uox} for the conventions. The Newman-Unti gauge \cite{Newman:1962cia} is adopted for the null basis and spin coefficients. In the retarded stereographic coordinates $(u,r,z,\bz)$, the null basis are chosen as
\be\begin{split}
&n=\frac{\p}{\p u} + U \frac{\p}{\p r} + X^A \frac{\p}{\p x^A},\\
&l=\frac{\p}{\p r},\label{tetrad}\\
&m=\omega\frac{\p}{\p r} + L^A \frac{\p}{\p x^A},
\end{split}\ee
where $A=(z,\bz)$ and $U,X^A,\omega,L^A$ are arbitrary functions of all coordinates. Correspondingly, the spin coefficients have the following simplifications,
\begin{align}
\pi=\kappa=\epsilon=0,\,\,\;\;\rho=\bar\rho,\;\;\,\,\tau=\bar\alpha+\beta.
\end{align}
The line element can be then written as
\begin{multline}
\td s^2 = -2 (U -\omega  \bomega ) \td u^2 + 2 \td u \td r \\
+g_{AB}\left[\td x^A + (L^A \bomega + \bL^A \omega -X^A)\td u \right]\left[\td x^B + (L^B \bomega + \bL^B \omega -X^B)\td u \right],
\end{multline}
where $g_{AB}=-L_A \bL_B - \bL_A L_B$ and $L_AL^A=0,\;L_A\bar L^A=-1$. The boundary conditions are chosen as  
\be\label{boundary}
\begin{split}
U=O(r),\quad \omega&=O(r),\quad \nu=O(r),\\
L^z=O(r),\quad X^A=O(r),& \quad \text{Im} [\mu]=O(r),\quad \lambda=O(r).
\end{split}
\ee
The only relaxation compared to \cite{Liu:2022uox} is that we remove the condition $\text{Im} [L^{\bz}]=O(r)$. It seems that the difference is minor. But the consequence is significant and very remarkable as we will show in the next pages. We postpone commenting on other relaxation for the gauge and boundary conditions after null boundary symmetries.

The gauge transformation in first order formalism is a combination of a diffeomorphism and a local Lorentz transformation. The gauge transformations of the tetrad and the spin connection are given by
\be
\begin{split}
&\delta_{\xi,\omega}{e_a}^\mu ={\xi}^\nu\partial_\nu
{e_a}^\mu-\p_\nu{\xi}^\mu{e_a}^\nu +{\omega_a}^b{e_b}^\mu, \\
&\delta_{\xi, \omega} \Gamma_{a b c} = {\xi}^\nu \partial_\nu \Gamma_{a b c} - e_c^\mu \p_\mu {\omega}_{a b} + {\omega_a}^{d}\Gamma_{dbc}+ {\omega_b}^{d}\Gamma_{adc} + {\omega_c}^{d}\Gamma_{abd} .
\end{split}
\ee
The residual gauge transformation that preserved the gauge and boundary conditions are explicitly determined by the symmetry parameters $f(u,z,\bz)$, $Y(z)$, $\bY(\bz)$ which generate a horizon diffeomorphism, and $\Lambda(u,z,\bz)$ which generates a complex Lorentz transformation. Note that $\Lambda(u,z,\bz)$ is purely imaginary, i.e., $\xbar\Lambda=-\Lambda$. For notational brevity, we will not involve $\xbar\Lambda$. But a minus sign should be understood when taking the complex conjugate of terms with $\Lambda$. The precise forms of the residual gauge transformations are
\be\begin{split}
&\xi^u=f,\quad\quad \xi^A=Y^A + \p_B f
  \int^r_0 \td r'[L^A \bL^B + \bL^AL^B ],\\\nn
&\xi^r=-\p_u f  r + \p_A
  f \int^r_0 \td r'[\omega \bL^A + \bomega L^A - X^A] ,
\end{split}
\ee
and
\be\begin{split}\nn
&\omega^{12}=\p_u f  + \p_A f X^A,\quad \omega^{13}=  - \p_A f \int^r_0
\td r'[\lambda L^A + \mu \bL^A],\\
& \omega^{23}= \bL^A \p_A f, \quad \omega^{34}=\Lambda  + \p_A f \int^r_0
  \td r'[(\bar{\alpha}-\beta) \bL^A - (\alpha - \bar{\beta}) L^A].
\end{split}
\ee
The constant order in $\xi^r$ is set to be zero by hand which fixes the $r=0$ null hypersurface to be the boundary. Alternatively, one can consider that the existence of a boundary at $r=0$ breaks the translational invariance along $r$ direction  \cite{Donnay:2015abr,Adami:2021nnf}.

By removing the conditions in the second line of \eqref{boundary}, one can obtain that $Y^z$ and $Y^{\bz}$ are arbitrary functions on the null boundary which will recover the full horizon diffeomorphism in \cite{Adami:2021nnf}. Correspondingly, such choice will significantly enhance the solution space. Since we mainly focus on new symmetries and charges in the present work, we will deal with the restricted case with independent Lorentz transformation which can not be seen from metric formalism. For the Lorentz part, one can relax some boundary conditions, i.e., only setting $U-\omega\bomega=O(r)$, to save the integration constants in $\omega^{13}$ and $\omega^{14}$. Such choice will enhance the solutions space and asymptotic symmetries. But we have checked that the enhanced symmetries do not arise any new charge. So we consider the enhanced symmetries trivial and the solution space is enhanced by trivial gauge transformation. Alternatively, one can absorb the enhanced fields from the solution space by changing of slicing at the charge level \cite{Adami:2020ugu,Ruzziconi:2020wrb,Adami:2021sko}. We just impose stronger boundary conditions here to eliminate those redundancies. One can also enhance the symmetries by removing the gauge condition $\tau=\xbar\alpha+\beta$. In such case, the null base vector $l$ is proportional to the gradient of a scalar field. Consequently, one can obtain an independent Lorentz transformation from $\omega^{12}$, namely replacing $\p_u f$ in $\omega^{12}$ and $\xi^r$ by an independent parameter $\omega^{12}_0$. Clearly, this new symmetry is a Weyl scaling of the null boundary as discussed in the metric formalism in \cite{Adami:2021nnf}. All in all, the only non-trivial Lorentz transformation in the current setup is from $\Lambda$. For simplicity, we will work in the simplest case but with full independent $\Lambda$ which is from the boundary conditions we choose in \eqref{boundary}.

If the symmetry parameters depend on the fields, one can use an adjusted Lie bracket which can subtract the changes in the symmetry transformation due to the variation of the fields \cite{Barnich:2010eb,Barnich:2011mi,Compere:2015knw}.  The adjusted bracket in the NP formalism is defined in \cite{Barnich:2019vzx}. The adjusted bracket is
\begin{equation}
  \begin{split}
& [\delta_{\xi_1,\omega_1},\delta_{\xi_2,\omega_2}]\phi^\alpha=
\delta_{\hat\xi,\hat\omega}\phi^\alpha,\\
& \hat\xi^\mu=[\xi_1,\xi_2]^\mu
-\delta_{\xi_1,\omega_1}{\xi}^\mu_2+\delta_{\xi_2,\omega_2}{\xi}^\mu_1,
  \\
 & {{(\hat{\omega})}_a}^b
 ={\xi_1}^\rho\p_\rho{\omega_{2a}}^b+{\omega_{1a}}^c{\omega_{2c}}^b
 -\delta_{\xi_1,\omega_1}{\omega_{2a}}^b
  -(1\leftrightarrow
  2),
\end{split}
\end{equation}
where $\phi^\alpha$ denotes an arbitrary field. The near horizon symmetry characterized by the parameters $(\xi[f,Y^A],\omega[\Lambda,f,Y^A])$
realize a symmetry algebra anywhere in the near horizon region with the adjusted bracket as
\begin{equation}\label{NHSG}
  \begin{split}
  &\hat \xi=\xi[\hat f,\hat Y^A],\quad \hat
  \omega=\omega[\hat\Lambda,\hat f,\hat Y^A],\\
  &\hat f=Y_1^A\p_A f_{2}+ f_{1}\p_u f_2 -(1\leftrightarrow
  2),\\
  & \hat Y^A=Y_1^B\p_B Y^A_2-Y_2^B\p_B Y^A_1,\\
  &\hat\Lambda=f_1 \p_u \Lambda_2 + Y_1^A \p_A \Lambda_2 - (1\leftrightarrow
  2).
\end{split}
\end{equation}
In particular, the symmetries form an algebra with the standard Lie bracket when $f$, $Y^A$ and $\Lambda$ are field independent on the horizon.

First order formalism is particularly adorable for computing near horizon boundary charges. Because the leading order charge only involves the leading order fields. All the leading order fields are integration constants which are free data. Hence it is not necessary to solve the radial NP equations for deriving the near horizon charges. For completeness, we work out the subleading order solutions of the radial equations which could be useful for computing the subleading near horizon charges. The solutions in near horizon expansion are
\bea
&&\Psi_0=\Psi_0^0  + \Psi_0^1 r + O(r^2),\label{psi0}\\
&&\rho=\rho_0 + (\rho_0^2 + \sigma_0 \xbar\sigma_0) r +   O(r^2),\label{rho}\\
&&\sigma=\sigma_0 + (2\rho_0 \sigma_0 + \Psi_0^0) r +  O(r^2),\label{sigma}\\
&&L^z= \bP \sigma_0 r +   O(r^2),\label{Lz}\\
&&L^{\bz}=P + P\rho_0 r  + O(r^2),\label{Lzb}\\
&&L_z=-\frac{1}{\bP} + \frac{\rho_0} {\bP} r + O(r^2),\\
&&L_{\bz}= \frac{\sigma_0}{P}r + O(r^2),\\
&&\alpha=\alpha_0 + (\alpha_0 \rho_0 + \beta_0 \xbar\sigma_0) r + O(r^2),\label{alpha}\\
&&\beta=\beta_0 + (\alpha_0 \sigma_0+\beta_0\rho_0 + \Psi_1^0) r + O(r^2),\label{beta}\\
&&\omega=-(\xbar\alpha_0 +\beta_0) r + O(r^2),\label{omega}\\
&&\tau=\tau_0+(\rho_0\tau_0 + \sigma_0\xbar\tau_0 + \Psi_1^0)r+O(r^2),\label{tau}\\
&&\Psi_1=\Psi_1^0 + \Psi_1^1 r  + O(r^2),\quad \Psi_1^1=4\rho_0 \Psi_1^0 + \xbar\eth \Psi_0^0,\label{psi1}\\
&& X^z=\bP \tau_0  r + O(r^2),\label{XA}\\
&&\mu=\mu_0 + (\mu_0\rho_0 + \Psi_2^0) r  + O(r^2),\quad \mu_0=\xbar\mu_0,\label{mu}\\
&&\lambda= \mu_0\xbar\sigma_0 r + O(r^2), \label{lambda}\\
&&\Psi_2=\Psi_2^0 + \Psi_2^1 r + O(r^2),\quad \Psi_2^1=  3\rho_0\Psi_2^0 + \xbar\eth\Psi_1^0,\label{psi2}\\
&&\gamma=\gamma_0 + (\alpha_0\tau_0 + \beta_0\xbar\tau_0 + \Psi_2^0) r + O(r^2),\label{gamma}\\
&& U=-(\gamma_0+\xbar\gamma_0) r  + O(r^2),\label{U}\\
&&\Psi_3=\Psi_3^0 + \Psi_3^1 r + O(r^2),\quad \Psi_3^1=  2\rho_0\Psi_3^0 + \xbar\eth \Psi_2^0, \label{psi3}\\
&&\nu=(\lambda_0\tau_0 +  \mu_0 \xbar\tau_0 + \Psi_3^0)r+O(r^2),\label{nu}\\
&&\Psi_4=\Psi_4^0 + \Psi_4^1 r +O(r^2), \quad \Psi_4^1= \rho_0\Psi_4^0 + \xbar\eth\Psi_3^0,  \label{psi4}
\eea
where quantities with subscript $0$ are integration constants of the radial differential equations. The ``$\eth$'' operator is defined as
\begin{equation}\begin{split}
&\eth \eta^s= P\p_{\bz} \eta^s + 2 s\xbar\alpha^0 \eta^s,\\
&\xbar\eth \eta^s=\bP \p_z \eta^s -2 s \alpha^0 \eta^s,\nn
\end{split}\end{equation}
where $s$ is the spin weight of the field $\eta$. The spin weights of relevant fields are listed in Table \ref{t1}.
\begin{table}[ht]
\caption{Spin weights}\label{t1}
\begin{center}\begin{tabular}{|c|c|c|c|c|c|c|c|c|c|c|c|c|c|c|c|c|c}
\hline
& $\eth$ & $\p_u$ & $\gamma^0$ & $\nu^0$ & $\mu^0$ & $\sigma^0$ & $\lambda^0$  & $\Psi^0_4$ &  $\Psi^0_3$ & $\Psi^0_2$ & $\Psi^0_1$ & $\Psi_0^0$   \\
\hline
s & $1$& $0$& $0$& $-1$& $0$& $2$& $-2$  &
  $-2 $&  $-1$ & $0$ & $1$ & $2$    \\
\hline
\end{tabular}\end{center}\end{table}

The integration constants are constrained as \footnote{Anther possibility is that $\mu_0=0$ and $\gamma_0=\xbar\gamma_0$ is an arbitrary function.}
\bea
&&\mu_0=-\frac12 \p_u \ln P\bP,\label{mu0}\\
&&\alpha_0=\frac12 (\xbar\tau_0  + \bP \p_z \ln P),\label{alpha0}\\
&&\beta_0= \frac12(\tau_0 - P \p_{\bz} \ln \bP) ,\label{beta0}\\
&&\gamma_0=\frac14 \p_u\ln \frac{P}{\bP}-\frac12 \mu_0 -\frac12 \p_u \ln \mu_0 ,\label{gamma0}\\
&&\Psi_4^0=0,\label{psi40}\\
&&\Psi_3^0=\xbar\eth\mu_0 + \mu_0 (\alpha_0 + \xbar\beta_0),\label{psi30}\\
&&\Psi_2^0=\mu_0 \rho_0 + \alpha_0 \xbar\alpha_0 + \beta_0 \xbar\beta_0 - 2 \alpha_0\beta_0 - P \p_{\bz} \alpha_0 + \bP \p_z \beta_0,\label{psi20}\\
&&\Psi_1^0=\xbar\alpha_0\rho_0 + \alpha_0\sigma_0 + \beta_0\rho_0 + \sigma_0 \xbar\beta_0 - \eth\rho_0 + \xbar\eth\sigma_0,\label{psi10}\\
&& \p_u \tau_0 = 2\eth\xbar\gamma_0 + 2 (\gamma_0 - \xbar\gamma_0 - \mu_0)\xbar\alpha_0 - 2\eth\mu_0 - 2\mu_0 \tau_0  -  \p_u(P \p_{\bz}\ln \bP), \label{tau0}\\
&& \p_u \rho_0 = \rho_0(\gamma_0+\xbar\gamma_0) - \mu_0\rho_0 + \xbar\eth \tau_0 - \Psi_2^0, \label{rho0}\\
&& \p_u\sigma_0=\eth\tau_0 - 2\tau^2_0 - \mu_0\sigma_0 + (3\gamma_0 - \xbar\gamma_0)\sigma_0. \label{sigma0}\\
&& \p_u \Psi_0^0 - 4\gamma_0 \Psi_0^0 + \mu_0 \Psi_0^0 = \eth \Psi_1^0 - 6\tau_0 \Psi_1^0 + 3\sigma_0 \Psi_2^0 \label{psi00}.
\eea
Our solution space is a subset of the one in \cite{Adami:2021nnf}. To recover our solution space, one just needs to set the fields in \cite{Adami:2021nnf} as
\be
{\cal U}^A\rightarrow0,\quad \eta\rightarrow 1,\quad  \Omega\rightarrow\frac{1}{P\bP},\quad \gamma_{z\bz}\rightarrow-1,\quad \gamma_{zz}\rightarrow0.
\ee
and
\be
\kappa\rightarrow - (\gamma_0+\xbar\gamma_0),\quad  \Upsilon^z\rightarrow \frac{2\tau_0}{P}.
\ee
Other simplifications of \cite{Adami:2021nnf} in this case are
\be\begin{split}
&{\cal D}_v=\p_v,\quad \Theta_l=\p_v\ln \Omega\rightarrow2\mu_0,\quad N^{AB}=0,\\
&\Gamma=-2\kappa+\p_v\ln\Omega+\p_v\ln \eta=2(\gamma_0+\xbar\gamma_0) + 2 \mu_0.
\end{split}
\ee
Those relations will be useful later to compare our charge with the one in \cite{Adami:2021nnf}.

Acting the residual gauge transformation on the near horizon fields yields their transformation laws as 
\begin{align}
&\delta_{\xi,\omega} \frac{1}{P} = f\p_u \frac{1}{P}  +Y^A \p_A \frac{1}{P}  +  \p_{\bz} Y^{\bz}  \frac{1}{P} - \Lambda  \frac{1}{P},\label{dP}\\
&\delta_{\xi,\omega} \ln P = f\p_u \ln P  +Y^A \p_A \ln P  -  \p_{\bz} Y^{\bz}  + \Lambda,\label{dlnP}\\
&\delta_{\xi,\omega} \frac{1}{P\bP} = f\p_u \frac{1}{P\bP}  +Y^A \p_A \frac{1}{P\bP}  +  \p_A Y^A  \frac{1}{P\bP},\label{dPPb}\\
&\delta_{\xi,\omega} \mu_0 =f\p_u \mu_0  + Y^A \p_A \mu_0  + \p_u f \mu_0 , \label{dmu}\\
&\delta_{\xi,\omega} \gamma_0 = f\p_u \gamma_0 + Y^A \p_A \gamma_0  + \p_u f \gamma_0  - \frac12 ( \p_u^2 f + \p_u \Lambda),\label{dgamma}\\
&\delta_{\xi,\omega} \tau_0= f\p_u \tau_0 + Y^A \p_A \tau_0 + \Lambda \tau_0 + 2 \eth  f  \gamma_0 -  \p_u(\eth f). \label{dalphabeta}\\
&\delta_{\xi,\omega} \frac{\tau_0}{P}= f\p_u \frac{\tau_0}{P} + Y^A \p_A \frac{\tau_0}{P} + \p_{\bz} \bY \frac{\tau_0}{P} +  2\p_{\bz} f  \gamma_0 - \p_u \ln P \p_{\bz} f - \p_u \p_{\bz} f. \label{dalphabetaP}\\
&\delta_{\xi,\omega} \rho_0 = f\p_u \rho_0  + Y^A \p_A \rho_0  - \p_u f \rho_0
 + \eth f \xbar\tau_0 + \xbar\eth f \tau_0 - P\bP\p_z\p_{\bz} f.\label{drho}\\
&\delta_{\xi,\omega} \sigma_0 = f\p_u \sigma_0  + Y^A \p_A \sigma_0 + 2 \Lambda \sigma_0 - \p_u f \sigma_0 + 3 \eth f \tau_0 - \eth^2 f.\label{dsigma}
\end{align}

\section{Near horizon charges from Palatini action}
\label{Palatinisection}

In this section, we will compute the boundary charge for the Palatini action. The Palatini Lagrangian in the form language is
\be\label{palatiniaction}
L_{Pa}=\frac{1}{32\pi G}\epsilon_{abcd} R^{ab}\wedge e^c \wedge e^d ,
\ee
where $R^{ab}=d \Gamma^{ab} + \Gamma^{ac}\wedge {\Gamma_c}^b$ is the curvature two form. The boundary charge from this Lagrangian is defined by\footnote{When the symmetry parameters are field independent, this expression will recover the one derived in \cite{Godazgar:2020kqd}.} 
\be\label{palatinicharge}
\sd {\cal H}_{Pa}= \frac{1}{32\pi G} \epsilon^{abcd} \int_{\partial \Sigma} \left[\delta (i_\xi \Gamma_{ab}  e_c \wedge e_d) - i_\xi( \delta \Gamma_{ab}\wedge e_c\wedge e_d) - \delta(\omega_{ab} e_c\wedge e_d)\right],
\ee
where $\partial \Sigma$ can be any constant-$u$ two surface on the horizon to evaluate the boundary charge. Inserting the solutions and the symmetry parameters yields
\begin{multline}\label{palatini}
\sd {\cal H}_{Pa}=\frac{1}{8\pi G}\int_{\partial \Sigma} \td z \td \bz \bigg[\delta\left(  \frac{\p_u f}{P\bP}-Y^z \frac{1}{P\bP} \frac{\xbar\tau_0}{\bP} - Y^{\bz} \frac{1}{P\bP} \frac{\tau_0}{P} \right)\\
- (\gamma_0+\xbar\gamma_0)\delta\frac{f}{P\bP} 
-f \left(\frac{2}{P\bP} \delta \mu_0 + \mu_0\delta\frac{1}{P\bP} \right) \bigg].
\end{multline}
This charge matches the one in \cite{Adami:2021nnf} explicitly using the relations in Section \ref{solution}. Since our charge only recovers a subset of the charge in \cite{Adami:2021nnf}, we will not repeat the charge algebra or the balance relations. Nevertheless, the main motivation of this section is to show that the Lorentz transformation characterized by the symmetry parameter $\Lambda$ does not have its associated charge from the Palatini action.

\section{Near horizon charges from Holst term}
\label{Holstsection}

The Holst term can be considered as the dual of the Palatini Lagrangian. The explicit form is 
\be
L_H=\frac{it}{16\pi G} R_{ab}\wedge e^a \wedge e^b,
\ee
where $t$ is the Holst term parameter. Though the Holst term is not a boundary term, it does not affect the equations of motion from the Palatini action \cite{Godazgar:2020kqd,DePaoli:2018erh}. The boundary charge derived from the Holst term is \cite{Godazgar:2020kqd}
\be
\sd {\cal H}_H= \frac{it}{16\pi G}  \int_{\partial \Sigma} \left[\delta(i_\xi \Gamma^{ab}  e_a \wedge e_b) - i_\xi( \delta \Gamma^{ab}\wedge  e_a \wedge e_b) -\delta( \omega^{ab} e_a\wedge e_b)\right].
\ee
Inserting the near horizon solutions and symmetries, the near horizon Holst charge reads 
\begin{multline}
\sd {\cal H}_H =\frac{it}{8\pi G}\int_{\partial \Sigma} \td z \td \bz \bigg\{ \delta\left[ \frac{Y^{\bz}}{P\bP}\left(\frac{ \xbar\alpha_0 - \beta_0}{P}\right) - \frac{Y^z}{P\bP}  \left(\frac{\alpha_0 - \xbar\beta_0}{\bP}\right)   - \frac{\Lambda}{P\bP}\right]\\
-(\gamma_0 - \xbar\gamma_0) \delta \frac{f}{P\bP} + f  \frac{\mu_0}{P^2\bP^2}( \bP \delta P- P \delta \bP)\bigg\}.
\end{multline}
Considering $f$ field independent and using the relations of the solution space, the charge can be rewritten as
\begin{multline}\label{holst}
\sd {\cal H}_H =\frac{it}{8\pi G}\int_{\partial \Sigma} \td z \td \bz \bigg[ \delta\left( \frac{Y^{\bz}}{P\bP}\p_{\bz}\ln \bP - \frac{Y^z}{P\bP}  \p_z \ln P  - \frac{\Lambda}{P\bP}\right)\\
 +\frac{f}{P\bP}\left(\p_u \ln P \delta\ln \bP - \p_u\ln \bP \delta \ln P \right)\bigg].
\end{multline}
Clearly, the Lorentz transformation has its own charge from the Holst term. Non-trivial boundary conditions bring boundary degrees of freedom which are labelled by boundary charges. Here, we provide a precise example that trivial terms which do not modify the local equations of motion will not only affect the boundary degrees of freedom via their modification to the boundary charges but will also increase new boundary degrees of freedom. Moreover, this new charge can not be seen from metric formalism. There will be one more charge than the generic analysis in metric formalism in four dimensions \cite{Adami:2021nnf}, namely five boundary charges on the three dimensional null boundary.

In general, the surface charge of a theory may not be integrable. One can use the Barnich-Troessaert prescription \cite{Barnich:2011mi} to split the charge into an integrable part and a flux part, 
\be
\sd {\cal H}_{(\xi,\omega)} =\delta{\cal H}^I_{(\xi,\omega)}  + {\cal F}_{(\xi,\omega)} (\delta \phi^\alpha;\phi^\alpha),
\ee
such that the surface charges satisfy the modified bracket\footnote{Note that the definition of the modified charge algebra is slightly different from the original proposal in \cite{Barnich:2011mi}. Because we used a different notation for the transformation law of the fields. Correspondingly, the generalized cocycle condition of the 2-cocycle term is also different.}
\be\label{bracket}
\begin{split}
&\delta_{(\xi_2,\omega_2)} {\cal H}^I_{(\xi_1,\omega_1)} + {\cal F}_{(\xi_2,\omega_2)} (\delta_{(\xi_1,\omega_1)} \phi^\alpha;\phi^\alpha):=\{ {\cal H}^I_{(\xi_1,\omega_1)},{\cal H}^I_{(\xi_2,\omega_2)} \}_{MB} \\
&\{ {\cal H}^I_{(\xi_1,\omega_1)},{\cal H}^I_{(\xi_2,\omega_2)} \}_{MB}={\cal H}^I_{(\hat{\xi},\hat{\omega})} + K_{(\xi_1,\omega_1),(\xi_2,\omega_2)},
\end{split}
\ee
where the parameters $(\hat{\xi},\hat{\omega})$ are defined from the adjusted bracket of the symmetries and $K_{(\xi_1,\omega_1),(\xi_2,\omega_2)}$ is the possible 2-cocycle term. We choose the integrable part of the Holst charge \eqref{holst} as
\be
{\cal H}_H^I=\frac{it}{8\pi G}\int_{\partial \Sigma} \td z \td \bz \left( \frac{Y^{\bz}}{P\bP}\p_{\bz}\ln \bP - \frac{Y^z}{P\bP}  \p_z \ln P  - \frac{\Lambda}{P\bP}\right), 
\ee
while the flux part as
\be
{\cal F}_H=\frac{it}{8\pi G}\int_{\partial \Sigma} \td z \td \bz \frac{f}{P\bP}\left(\p_u \ln P \delta\ln \bP - \p_u\ln \bP \delta \ln P \right).
\ee
The modified algebra yields the following 2-cocycle term
\be\label{K}
{K_H}_{(\xi_1,\omega_1),(\xi_2,\omega_2)}=\frac{it}{8\pi G}\int_{\partial \Sigma} \td z \td \bz \frac{1}{P\bP}\left(f_1 \p_u \Lambda_2 - f_2 \p_u \Lambda_1\right).
\ee
A total derivative term $\p_A \cY^A$ has been dropped when verifying the charge algebra where
\be\begin{split}
&\cY^z=i\left(Y_1^{\bz} \p_{\bz} \ln \bP - Y_1^z \p_z \ln P\right)\frac{Y^z_2}{P\bP}-i\Lambda_1 \frac{Y_2^z}{P\bP}-i\frac{f_2 \p_u \ln P}{P\bP}Y_1^z ,\\
&\cY^{\bz}=i\left(Y_1^{\bz} \p_{\bz} \ln \bP - Y_1^z \p_z \ln P\right)\frac{Y^{\bz}_2}{P\bP}-i\Lambda_1 \frac{Y_2^{\bz}}{P\bP}+i\frac{f_2 \p_u \ln \bP}{P\bP}Y_1^{\bz} ,
\end{split}\ee
and the imaginary unit is from the coupling. One can prove that the 2-cocycle term \eqref{K} satisfies the suitably generalized cocycle condition
\be
{K_H}_{[(\xi_1,\omega_1),(\xi_2,\omega_2)],(\xi_3,\omega_3)}+\delta_3 {K_H}_{(\xi_1,\omega_1),(\xi_2,\omega_2)} + \text{cyclic}(1,2,3)=0,
\ee
up to a total derivative term
\be
\p_A \left(i \frac{1}{P\bP} f_1 \p_u \Lambda_2 Y_3^A - i \frac{1}{P\bP} f_3 \p_u \Lambda_2 Y_1^A \right) + \text{cyclic}(1,2,3).
\ee
The boundary charge algebra yields the balance equation for the Holst charges as
\be\label{balance}
\begin{split}
\frac{\p}{\p u} {\cal H}_H^I&=\delta_{\frac{\p}{\p u}} {\cal H}_H^I+{{\cal H}_H^I}_{(\p_u \xi, \p_u \Lambda)}=-{{\cal F}_H}_{\frac{\p}{\p u}} (\delta_{(\xi,\omega)} \phi^\alpha;\phi^\alpha) + {K_H}_{(\xi,\omega),(\frac{\p}{\p u},0)},
\end{split}
\ee
which can be verified directly using \eqref{dlnP} and its complex conjugate. The non-conservation of the charges is induced by the flux going through the horizon which is characterized by the unconstrained $u$-dependence of $P$ and $\bP$.

\section{Concluding remarks}

In this paper, we study the near horizon boundary charges in the NP formalism. We find that there is one new charge from the Holst term that is associated to the Lorentz transformation. The new charge can not be seen from the metric formalism. Hence, we have formulated a boundary system that has one more charge than the one in \cite{Adami:2021nnf}. Interestingly, the contribution to the boundary charge from the Holst term can be equivalently obtained from the freedom in the definition of boundary charges \cite{Oliveri:2020xls} in the covariant phase space formalism \cite{Lee:1990nz,Iyer:1994ys}. More precisely, the symplectic potential from the Holst term is
\be
\theta_H=\frac{it}{16\pi G} \delta\Gamma^{ab}\wedge e_a\wedge e_b.
\ee
On-shell (vanishing of the torsion), this symplectic potential can be written as
\be
\theta_H=\frac{it}{16\pi G} \td \left(\delta e_a\wedge e^a\right),
\ee
which is precisely a Y-freedom in the definition of symplectic form. Hence, we have shown a precise example that whether or not a residual gauge transformation is trivial is, to our surprise, relevant to the freedom in the definition of the boundary charges. Alternatively, this can provide a criterion to fix the freedom in the definition of the boundary charges. 

There are several extensions and applications of our results. We mention some of them for future directions.
\begin{itemize}
\item 
A similar analysis can be performed on a casual boundary. Naturally, one would expect that there should be more charges derived in that case \cite{Adami:2022ktn} in particular from Lorentz transformations.
\item
Boundary degrees of freedom from null hypersurface admit a thermodynamical description  \cite{Adami:2021kvx}. What is the thermodynamical description of the new Lorentz charge is a very interesting question that needs to be addressed elsewhere.
\item
The 2-cocycle term of the charge algebra has horizon area dependence. It may be relevant to the investigation of the microscopic origin of black hole entropy following the line of \cite{Strominger:1997eq,Guica:2008mu}.
\end{itemize}

\section*{Acknowledgments}

The authors thank Hai-Shan Liu for the early collaboration on this project. The authors thank Mahdi Godazgar, Hai-Shan Liu, Roberto Oliveri and Shahin Sheikh-Jabbari for valuable discussions. This work is supported in part by the National Natural Science Foundation of China under Grant No. 11905156 and No. 11935009.

\bibliography{ref}

\end{document}